\documentclass[10pt,a4paper,english]{endm}
\usepackage[T1]{fontenc}
\usepackage[latin9]{inputenc}
\usepackage{amstext}
\usepackage{amssymb}
\usepackage{stmaryrd}
\usepackage{graphicx}

\makeatletter

\usepackage{endmmacro}
\usepackage{graphicx}

\usepackage{tikz,tkz-tab,tkz-graph}
\usepackage{relsize}
\usepackage{balance}
\usepackage{faktor}

\usetikzlibrary{arrows,automata}
\usetikzlibrary{positioning, fadings}
\tikzset{
    state/.style={
           rectangle,
  fill=#1!5!white,
           draw=#1, very thick,
           minimum height=2em,
           inner sep=2pt,
           text centered,
           },
    highlight/.style={
           rectangle,
  fill=#1!50!white,
           rounded corners,
           draw=#1, very thick,
           minimum height=2em,
           inner sep=2pt,
           text centered,
           },
coeff/.style={
           circle,
           draw=black, very thick,
           minimum height=2em,
           inner sep=2pt,
           text centered,
           },	
}

\tikzset{
    invisible/.style={opacity=0},
    visible on/.style={alt=#1{}{invisible}},
    alt/.code args={<#1>#2#3}{%
      \alt<#1>{\pgfkeysalso{#2}}{\pgfkeysalso{#3}} 
    },
  }

\tikzset{fontscale/.style = {font=\relsize{#1}}
    }

\AtBeginDocument{

}

\makeatother

\usepackage{babel}
\begin{document}
\begin{verbatim}\end{verbatim}\vspace{2.5cm}
\begin{frontmatter}
\title{Hypergraph Modeling and Visualisation of Complex Co-occurence Networks}
\author[CERN,Unige]{X. Ouvrard\thanksref{SupportCERN}},
\author[CERN]{J.M. Le Goff} and 
\author[Unige]{S. Marchand-Maillet}
\address[CERN]{CERN, CH-1211 Geneva 23}
\address[Unige]{University of Geneva, CUI, 7 route de Drize, Battelle   A, CH-1227 Carouge}
\thanks[SupportCERN]{This document is part of X. Ouvrard article work supervised by Pr. S. Marchand-Maillet and J.M. Le Goff and founded by a doctoral position at CERN.}
\thanks[XOMail]{Email: \href{mailto:xavier.ouvrard@cern.ch}
{\texttt{\normalshape    xavier.ouvrard@cern.ch}}}
\begin{abstract}
Finding inherent or processed links within a dataset allows to discover potential knowledge. The main contribution of this article is to define a global framework that enables optimal knowledge discovery by visually rendering co-occurences (i.e. groups of linked data instances attached to a metadata reference) - either inherently present or processed - from a dataset as facets. Hypergraphs are well suited for modeling co-occurences since they support multi-adicity whereas graphs only support pairwise relationships. This article introduces an efficient navigation between different facets of an information space based on hypergraph modelisation and visualisation.
\end{abstract}
\end{frontmatter}

\section{Introduction}

Having insight into non-numerical data calls for the gathering of
instances: classically (multi-entry) frequency arrays of occurences
are used. To get further insight into data instances of a given type,
one can regroup them using their links to instances of another type
- used as reference. It generates a family of co-occurence sets that
can be viewed as a facet of the information space. Navigating accross
the different facets is achieved by iterating this process between
different types of interest while keeping the same reference type:
any of these types can be used as a reference. We use a publication
dataset as breadcrumb trail example.

Previous approaches using a reference to articulate the different
facets of an information space exist \cite{dork12PivotPaths:StrollingthroughFacetedInformationSpaces,zhao2013interactive,hadlak2015survey}.
\cite{agocs2017interactive} proposes a graph-based framework which
provides insights into the different facets of an information space
based on user-selected perspectives, combining type of reference and
of co-occurences. \cite{taramasco2010academic} shows how the keeping
of $m$-adic relationships can help in gaining understanding in the
network evolution.

This article provides a hypergraph-based framework that supports interactions
between the different facets of an information space for optimal knowledge
discovery. The dataset - mostly textual - refers to physical entities
with unique individual references. Data instances are attached to
metadata instances. We suppose that there is no metadata instance
that doesn't have a data instance attached to it.

\section{Modeling co-occurences in datasets}

\label{sec:Hypergraph-modeling}

Hypergraphs suits well the storage of co-occurence information with
references. A \textbf{hypergraph} $\mathcal{H}=\left(V,E\right)$
is a hyperedge family $E=\left\{ e_{i}:e_{i}\subseteq V\land i\in\left\llbracket p\right\rrbracket \right\} $\footnote{$\left\llbracket k;n\right\rrbracket $ corresponds to $\left\{ i:i\in\mathbb{N}\land k\leqslant i\leqslant n\right\} $
and $\left\llbracket n\right\rrbracket $ to $\left\llbracket 1,n\right\rrbracket $.} over the vertex set $V=\left\{ v_{i}:i\in\left\llbracket n\right\rrbracket \right\} $
\cite{bretto2013hypergraph}. A hypergraph where the hyperedges are
distinct one-to-one is said \textbf{with no repeated hyperedge}. In
\cite{stell2012relations}, a hypergraph is a triple $\mathcal{H}=\left(V,E,i\right)$
with $V$ a vertex set, $E$ a hyperedge set and $i\,:\,E\rightarrow\mathcal{P}\left(V\right)$\footnote{ $\mathcal{P}\left(V\right)$ is the power set of $V$}
an incidence function. Considering a map $w:E\rightarrow\mathbb{R}^{++}$
the hypergraph $\mathcal{H}_{w}=\left(V,E,w\right)$ - or $\mathcal{H}_{w}=\left(V,E,i,w\right)$
- is said \textbf{weighted}.

\subsection{Allowing navigation}

Relational database schema are hypergraphs of metadata instances where
the hyperedges gather table metadata: normalized forms are linked
to the properties of the hypergraphs modeling them \cite{fagin1983degrees}.
In graph databases, the schema\footnote{although not required \cite{McColl:2014}}
represents the relationships between the vertex types. The \textbf{schema
hypergraph }$\mathcal{H}_{\text{Sch}}=\left(V_{\text{Sch}},E_{\text{Sch}},i_{\text{Sch}}\right)$
represents these relationships as hyperedges.

Each data instance stored in the dataset is labeled using a labeling
function on the vertices of $V_{\text{Sch}}$. Hyperedges of the schema
itself can be labeled by another labeling function over another label
set.

Types of visual or referencing interest are selected in a subset $U$
of $V_{\text{Sch}}$ to generate $\mathcal{H}_{X}=\left(V_{X},E_{X},i_{X}\right)$
the \textbf{extracted schema hypergraph} where $V_{X}=U$, $E_{X}=\left\{ e\cap U:e\in E_{\text{Sch}}\right\} $
and $i_{X}=\left.i_{\text{Sch}}\right|_{E_{X}}$ .

From $\mathcal{H}_{X}$, we build the \textbf{reachability hypergraph
}$\mathcal{H}_{R}=\left(V_{R},E_{R},i_{R}\right)$ with $V_{R}=V_{X}$
as vertex set, the hyperedges of $\mathcal{H}_{R}$ are the connected
components $E_{\text{cc}}\left(\subset E_{X}\right)$ of $\mathcal{H}_{X}$
- regrouped in $C_{X}$, the set of connected components of $\mathcal{H}_{X}$
- and $\forall e_{R}\in E_{R}:i_{R}\left(e_{R}\right)=\bigcup\limits _{E_{\text{CC}}\in C_{X}}\bigcup\limits _{e\in E_{\text{CC}}}i_{X}(e)$.

Last at the level of metadata, the \textbf{navigation hypergraph}
is built by choosing a nonempty subset of possible reference vertices
$R_{\text{ref}}$ in a hyperedge $e_{R}\in E_{R}.$ Non empty subsets
of $R_{\text{ref}}$ allow to generate possible hyperedges of the
navigation hypergraph $\mathcal{H}_{N}=\left(V_{N},E_{N}\right)$
where $V_{N}=e_{R},$ $E_{N}=\left\{ e_{R}\backslash R:R\subseteq R_{\text{ref}}\land R\neq\emptyset\right\} .$
Navigation is possible without changing reference inside a hyperedge
of $\mathcal{H}_{N}$.

In a publication dataset, typical metadata is: \textit{publication
id}, title, abstract, authors, affiliations, addresses, \textit{author
keyword}s, \textit{subject categories},\textit{ countries}, \textit{organisations},...\footnote{Metadata of interest for visualisation or referencing are in italic}
A possible navigation hyperedge is: \{author keywords, organisations,
country, subject category\} with publication id as reference.

\subsection{Facet visualisation hypergraphs}

Each physical entity $d$ in a dataset $\mathcal{D}$ is described
by a unique physical reference $r$ and a set of data instances of
different types $\alpha\in V_{\text{Sch}}$. The types are obtained
from the metadata - for instance in publications: organisation, author
keywords, country. We write $A_{\alpha,r}=\left\{ a_{1},...,a_{\alpha_{r}}\right\} $
the set of values of type $\alpha$ that are attached to $d.$ $A_{\alpha,r}$
is possibly the emptyset if no value of type $\alpha$ is attached
to $d$. Hence $d$ is fully described by: $\left(r,\left\{ A_{\alpha,r}:\alpha\in V_{\text{Sch}}\right\} \right).$ 

In the navigation hypergraph, each hyperedge $e_{N}\in E_{N}$ describes
accessible facets relatively to a reference type. A facet will show
co-occurences of a chosen type $\alpha\in e_{N}$ built relatively
to reference instances of type $\rho\in V_{N}\backslash e_{N}$ ($\alpha/\rho$
as short). For example, in a publication dataset, with organisation
as reference, one can retrieve all subject categories that are common
to a given organisation.

Performing a search on the dataset will retrieve a set $\mathcal{S}$
of physical references $r$. A facet will be represented by the visualisation
hypergraph of co-occurences of type $\alpha/\rho$. The set of all
values of type $\rho$ is defined by $\mathcal{\Sigma}_{\rho}=\bigcup\limits _{r\in\mathcal{S}}A_{\rho,r}$.
Each value of type $\rho$ is mapped to a set of physical references
in which they appear, using $r_{\rho}:v\in\Sigma_{\rho}\mapsto R_{v}$
where $R_{v}=\left\{ r:v\in A_{\rho,r}\right\} $. The set of values
of type $\alpha$ relatively to the reference $v$ is $\bigcup\limits _{r\in R_{v}}A_{\alpha,r}=e_{\alpha,v}.$ 

Hence the \textbf{raw visualisation hypergraph} for the facet of type
$\alpha/\rho$ attached to the search $\mathcal{S}$ is $\mathcal{H}_{\alpha/\rho,\mathcal{S}}=\left(\bigcup\limits _{r\in\mathcal{S}}A_{\alpha,r},\left(e_{\alpha,v}\right)_{v\in S_{\rho}}\right)$.

Some hyperedges can possibly point to the same subset of vertices.
In this case, we build a reduced visualisation weighted hypergraph
from the raw visualisation hypergraph. We define: $g_{\alpha}:v\mapsto e_{\alpha,v}$
and $\mathcal{R}$ the equivalence relation such that: $\forall v_{1}\in\mathcal{\Sigma_{\rho}}$,
$\forall v_{2}\in\Sigma_{\rho}$: $v_{1}\mathcal{R}v_{2}\Leftrightarrow g_{\alpha}\left(v_{1}\right)=g_{\alpha}\left(v_{2}\right).$ 

Considering $\overline{v}\in\mathcal{\Sigma_{\rho}}\big/\mathcal{R}$\footnote{$\mathcal{\Sigma_{\rho}}\big/\mathcal{R}$ is the quotient set of
$\mathcal{\Sigma_{\rho}}$ by $\mathcal{R}$}, we write $\overline{e_{\alpha,\overline{v}}}=g_{\alpha}\left(v\right)$
where $v\in\overline{v}$.

$\text{\ensuremath{\overline{E_{\alpha}}}}=\left\{ \overline{e_{\alpha,\overline{v}}}:\overline{v}\in\mathcal{\Sigma_{\rho}}\big/\mathcal{R}\right\} $
is the support set of the multiset\footnote{In a multiset repetitions of elements are allowed. For further details
\cite{radoaca2015properties}.} $\left\{ \left\{ e_{\alpha,v}:v\in\mathcal{\Sigma_{\rho}}\right\} \right\} $:
$\overline{e_{\alpha,\overline{v}}}\in\overline{E_{\alpha}}$ is of
multiplicity $w_{\alpha}\left(\overline{e_{\alpha,\overline{v}}}\right)=\left|\overline{v}\right|$
in this multiset. 

It yields: $\left\{ \left\{ e_{\alpha,v}:v\in\mathcal{\Sigma_{\rho}}\right\} \right\} =\left\{ \overline{e_{\alpha,\overline{v}}}^{w_{\alpha}\left(\overline{e_{\alpha,\overline{v}}}\right)}:\overline{v}\in\mathcal{S_{\rho}}\big/\mathcal{R}\right\} $

Let $\tilde{g_{\alpha}}:\overline{v}\in\mathcal{\Sigma_{\rho}}\big/\mathcal{R}\mapsto e\in\overline{E_{\alpha}}$,
then $\tilde{g_{\alpha}}$ is bijective. $\tilde{g_{\alpha}}^{-1}$
allows to retrieve the class associated to a given hyperedge; hence
the associated values of $\mathcal{\Sigma_{\rho}}$ to this class
- which will be important for navigation. The references associated
to $e\in\overline{E_{\alpha}}$ are $\bigcup\limits _{v\in\tilde{g_{\alpha}}^{-1}(e)}r_{\rho}\left(v\right).$
The \textbf{reduced visualisation weighted hypergraph} for the search
$\mathcal{S}$ is defined as $\mathcal{H}_{\alpha/\rho,w_{\alpha},\mathcal{S}}=\left(\bigcup\limits _{r\in\mathcal{S}}A_{\alpha,r},\overline{E_{\alpha}},w_{\alpha}\right).$

\subsection{Navigability through facets}

Keeping the same search $\mathcal{S}$ and reference $\rho$, the
sets $R_{v},v\in\mathcal{\Sigma_{\rho}}$ remain the same between
the different facets: considering another type $\alpha'\in e_{N}$
and using the same reference $\rho$, another visualisation hypergraph
$\mathcal{H}_{\alpha'/\rho}$ is built.

Let $\alpha$ being the current type and $\mathcal{H}_{\alpha/\rho,w_{\alpha}}$
being the current visualisation hypergraph. Focusing on a subset of
vertices $A\subseteq A_{\alpha,S}$, we retrieve the corresponding
hyperedge subset $\left.\overline{E_{\alpha}}\right|_{A}=\left\{ e:e\in\overline{E_{\alpha}}\land\left(\exists x\in e:x\in A\right)\right\} $
of $\overline{E_{\alpha}}$ which contains at least one element of
$A.$ Using $\tilde{g_{\alpha}}^{-1}$ we get for each $e\in\left.\overline{E_{\alpha}}\right|_{A}$
the class associated to the hyperedge $\overline{v}$, building the
set $\left.\overline{V}\right|_{A}=\left\{ \tilde{g_{\alpha}}^{-1}(e):e\in\left.\overline{E_{\alpha}}\right|_{A}\right\} .$
The references of type $\rho$ used to build the co-occurences are:
$\mathcal{V}_{\rho,A}=\left\{ v:\forall\overline{v}\in\left.\overline{V}\right|_{A}:v\in\overline{v}\right\} $.
From each element $v$ of $\mathcal{V}_{A}$, the set of physical
references $R_{v}$ is retrieved, considering $\left.r_{\rho}\right|_{\mathcal{V}_{\rho,A}}$
as the restriction of $r_{\rho}$ to $\mathcal{V}_{\rho,A}.$ It yields
to the physical reference set: $\mathcal{S}_{A}=\bigcup\limits _{v\in\nu_{\rho,A}}R_{v}.$

From these physical references, one can switch to another facet of
the same search with the same reference type $\rho$. Let $\alpha'$
be the targetted type. Then only $\left.\mathcal{H}_{\alpha'/\rho}\right|_{A}=\left(\bigcup\limits _{r\in\mathcal{S}_{A}}A_{\alpha',r},\left(e_{\alpha',v}\right)_{v\in\mathcal{V}_{\rho,A}}\right)$
will be processed as raw visualisation hypergraph, using $\mathcal{S}_{A}$
as reference search set in the former paragraph. To obtain the related
reduced weighted version we use the same approach as above. The set
of co-occurences retrieved include all occurences that have co-occured
with one of the element selected in the first facet.

Of course if $A=A_{\alpha,S}$ the reduced visualisation hypergraph
will contain all the instances of type $\alpha'$ attached to physical
entities of the search $\mathcal{S}$.

Ultimately, by building a multi-dimensional network organised around
types, one can retrieve very valuable information from combined data
sources. This process can be extended to any number of data sources
as long as they share one or more types. Otherwise the reachability
hypergraph is not connected and only separated navigations will be
possible.

\section{Conclusion}

\label{subsec:The-DataEdre}

\begin{figure}
\centering \includegraphics[scale=0.2]{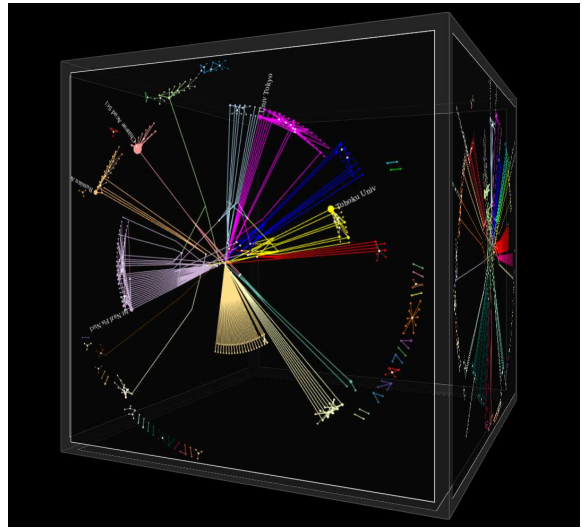}\caption{The DataHedron.}
\label{Fig: DataEdre}
\end{figure}

Using the connected components of the extracted schema we have enabled
the possibility of navigating the dataset. An application of the hypergraph
modeling framework is the DataHedron shown in Figure \ref{Fig: DataEdre}:
it enables easy navigation between facets of the information space.
It is a 2.5D representation of the information space where each DataHedron
face embeds a visualisation hypergraph. Navigation through facets
is articulated via the references that links one facet with another.
The link by references is realised on one face of the DataHedron.
Combining this framework with search tools allows to have deep insight
into a dataset.

{\small{}\bibliographystyle{elsarticle-num}
\addcontentsline{toc}{chapter}{\bibname}\bibliography{/home/xo/cernbox/these/000-thesis_corpus/biblio/references}
}{\small\par}
\end{document}